\documentclass[epj]{svjour}
\usepackage{graphics}
\begin{document}

\title{Polarized electric dipole moment of well-deformed reflection asymmetric nuclei}

\author{V. Yu. Denisov}
\institute{
Institute for Nuclear Research, Prospect Nauki 47,
03680 Kiev, Ukraine }%

\date{\today}

\abstract{
The expression for polarized electric dipole moment of well-deformed reflection asymmetric nuclei is obtained in the framework of liquid-drop model in the case of geometrically similar proton and neutron surfaces. The expression for polarized electric dipole moment consists of the first and second orders terms. It is shown that the second-order correction terms of the polarized electric dipole moment are important for well-deformed nuclei.
\PACS{
   {23.20.-g}{Electromagnetic transitions}  \and
   {23.20.Lv}{	Gamma-transitions and level energies}
   } 
} 
\maketitle

\section{Introduction}

Reflection asymmetric deformation of nucleus induces the proton-neutron redistribution. As a result, the proton or neutron density distributions became slightly polarized and reflection asymmetric in the nuclear volume. Due to such density polarization the position of proton center of mass is shifted relatively the nuclear center of mass; therefore reflection asymmetric nuclei have the polarized electric dipole moment (PEDM).

The PEDM of nuclei with quadrupole and octupole surface deformations was firstly obtained by V. M. Strutinsky in 1956 \cite{strutinsky} in the framework of liquid-drop model. A short time later A. Bohr and B. R. Mottelson was evaluated the PEDM in the same model \cite{b-m1}, but Strutinsky's derivation is the correct one \cite{b-m2}. The PEDM was found for non-axial nuclei with quadrupole and octupole deformations in Refs. \cite{lipas,leper}. Note the PEDM discussed in Refs. \cite{strutinsky,b-m1,b-m2,lipas,leper} is only related to the proton-neutron polarization in the volume of nuclei with quadrupole and octupole surface deformations.

However the proton-neutron density polarization in the nuclear volume induces the variation of proton and neutron radii and, therefore, leads to the corresponding surface contribution into the PEDM. The expression for PEDM with volume and surface contributions was derived in Ref. \cite{DMS} in the framework of the droplet model for axial nuclei with the proton radius $R_p(\theta)$ in the form
\begin{equation}
\frac{R_p(\theta)}{R_{0p}} = F(\theta) = \left[1+\sum_{\ell=2}^L \beta_\ell Y_{\ell 0}(\theta)\right].
\end{equation}
Here $R_{0p}$ is the proton radius of spherical nucleus, $\beta_\ell$ is the deformation parameter and $Y_{\ell 0}(\theta)$ is the spherical harmonic function. Both spherical and deformed nuclei have a neutron skin of constant thickness in the framework of the droplet model, see Ref. \cite{DMS} and papers cited therein, therefore the neutron radius $R_n(\theta)$ of deformed nuclei is not proportional the proton one $R_p(\theta)$ in the droplet model. Due to this expression for the PEDM obtained in the droplet model consists of the volume and surface charge redistribution contributions as well as the contribution related to the neutron skin thickness \cite{DMS,MS}. The neutron skin thickness contribution arises precisely from the non-coincidence of the centers of mass of a uniform skin and of the volume it encloses \cite{DMS,MS}.

The expression for PEDM for nuclei with geometrically similar proton and neutron surfaces, i.e. when proton and neutron radii have the same angular dependence
\begin{equation}
\frac{R_p(\theta)}{R_{0p}} = \frac{R_n(\theta)}{R_{0n}} = F(\theta),
\end{equation}
was obtained in Ref. \cite{D89}. Here $R_{0n}$ is the neutron radius of spherical nucleus. The neutron skin thickness depends on $\theta$, when radii of the proton and neutron surfaces are proportionally to each other. The PEDM consists of the volume and surface charge redistribution contributions only in this case \cite{D89}, because the neutron skin thickness contribution equals zero for geometrically similar proton and neutron surfaces, see for details \cite{MS,D89,D92} and Sec. II. The neutron skin center of mass coincides with the nucleus the ones \cite{D89,D92}.

The expressions for volume and surface parts of the PEDM for non-axial nuclei with arbitrary multipole deformations and geometrically similar proton and neutron surfaces are given in Ref. \cite{DD96}. The equilibrium shapes of some nuclei are non-axial reflection asymmetric \cite{Skalski91}. Moreover PEDM can be arisen at non-axial reflection asymmetric surface vibrations \cite{DD96}.

We emphasize that the PEDM obtained in the first non-zero order on multipole deformations of nuclear surface is proportional to $\beta_\ell \beta_{\ell+1}$ and all expressions for the PEDM presented in Refs. \cite{strutinsky,b-m1,b-m2,lipas,leper,DMS,MS,D89,D92,DD96} are derived in this approximation.

Numerical study of the PEDM in well-deformed nuclei in Ref. \cite{Skalski} shows that the first approximation for PEDM is strongly underestimated the numerical one. Moreover the difference between the numerical and first-order values of PEDM increases with values of deformation parameters strongly \cite{Skalski}.

The values of PEDM have been also evaluated in the frameworks of various semi-microscopic or microscopic approaches, see Refs. \cite{D92,Skalski,BN-RMP,BN-D,TsKN,BMSV} and papers cited therein.

The nuclei with quadrupole and octupole deformations, $E1$ transitions and the PEDM are studied intensively recently \cite{DDa,DDb,DDc,Minkov,RRF,expDipMom1a,expDipMom1b,expDipMom2,th227-231a,th227-231b}. The PEDM plays important role in various phenomena of well-deformed reflection asymmetric nuclei. Thus Karpeshin has shown that well-deformed fission fragments of such shapes formed during prompt fission give rise to both the anomalous $E1$ internal conversion \cite{Karpeshin92} and the prompt gamma radiation \cite{Karpeshin10,Karpeshin10a} related to the PEDM. The left-right asymmetry of fission induced by polarized neutrons can be also linked to the PEDM \cite{Karpeshin08}. The $E1$ transitions possibly linked to octupole vibrations around super-deformed shape can be also enhanced by the PEDM \cite{SD-oct1,SD-oct2,SD-oct3,SD-oct4,SD-oct5}. Strong $E1$ transitions related to the low-energy shape oscillations of negative parity in the first and second (isomeric) minima in actinides are also connected to the PEDM \cite{KS}.

However application of expression for the PEDM obtained in the first order for well-deformed nuclei is questionable as pointed by Skalski \cite{Skalski}. Therefore it is desirable to obtain expression for the PEDM in the next order at least, which is the second order approximation for the PEDM contained terms proportional to $\beta_\ell \beta_{\ell^\prime} \beta_{\ell^{\prime \prime}}$. Such expression should be helpful and practical for description of various nature $E1$ transition in well-deformed nuclei.

The PEDM evaluated in the framework semi-microscopic approaches \cite{BN-RMP,BN-D} consists of macroscopic and microscopic contributions. The microscopic contribution of PEDM is evaluated without applying the perturbation approach on the surface deformation parameters, while macroscopic the one is evaluated by using expressions obtained in the first non-zero order on multipole deformations. Therefore more accurate expression for the macroscopic part of the PEDM improves the accuracy of the PEDM evaluated in the framework of semi-microscopic models.

It is well-known that shell effects are reduced in heated nuclei. Due to this expression for PEDM has reliable accuracy for highly excited fission fragments. The shape of fission fragments after rupture has appreciable reflection asymmetry. Microscopic calculation of PEDM in heated nuclei has not done up to now. Therefore expression for the PEDM obtained in the next order at least can be useful for evaluation of various effects related to dipole moment of fission fragments.

Various proposals on neutron skin are discussed recently. The neutron skin of permanent thickness in deformed nuclei is widely applied in the framework of the droplet model \cite{DMS,MS} and other papers related to this model. The neutron skin with thickness depended on $\theta$ and related to well-known and widely-used relation for the proton and neutron radii (2) is also very common in nuclear physics; see, for example, the description of the proton and neutron mean field radii and etc. \cite{D92,Skalski,BN-D,thskin1,thskin2,thskin3,thskin4,thskin5,thskin6,thskin7,thskin8}. We emphasize that the shapes of potential and density distributions are geometrically similar due to a consistency between the density distribution and the corresponding mean-field potential \cite{D92}. More complex approximation to the neutron skin thickness related to different neutron versus proton deformations of nuclei is discussed in Ref. \cite{pomorski1,pomorski2}. The thickness of neutron skin has complex angular dependence on $\theta$ in such case.

Diverse approximations for neutron skin shape are reasonable for small and medium deformed nuclei close to the beta-stability line. The ratio (2) between the proton and neutron radii is, probably, discussable for extremely-deformed and/or neutron-rich nuclei as pointed in Ref. \cite{MS}. Current experimental studies of neutron distribution on the surface of nuclei devote to the neutron skin thickness in spherical nuclei mainly, see Refs. \cite{skin1,skin2} and papers cited therein. Available experimental data \cite{skin1,skin2} cannot support firmly any of these approximations on the neutron skin thickness in deformed nuclei.

Relationship between proton and neutron surfaces or potential described by Eq. (2) is very widely used in nuclear physics for small, medium and even well-deformed nuclei \cite{D92,Skalski,BN-D,thskin1,thskin2,thskin3,thskin4,thskin5,thskin6,thskin7,thskin8}. Therefore, in Sec. II we derive expressions for the PEDM for the geometrically similar proton and neutron surfaces, which takes into account the first ($\propto \beta_\ell \beta_{\ell+1}$) and second ($\propto \beta_\ell \beta_{\ell^\prime} \beta_{\ell^{\prime \prime}}$) orders contributions. Discussion of obtained expressions, numerical results and conclusion are given in Sec. III.

\section{Model and expression for PEDM}

Let us consider the axial nucleus with proton and neutron radii described by Eq. (2). There are no any density polarizations in spherical nuclei, therefore the equilibrium neutron and proton density distributions in deformed nucleus can be presented as $\rho_n=\rho_{0n}+\delta\rho_n$ and $\rho_p=\rho_{0p}+\delta\rho_p$. Here $\rho_{0n}=3N/(4\pi R_{0n}^3)$ and $\rho_{0p}=3Z/(4\pi R_{0p}^3)$ are the equilibrium neutron and proton densities in spherical nucleus, $\delta\rho_n$ and $ \delta\rho_p$ are the variations of neutron and proton densities induced by surface deformation, $Z$ and $N$ are the numbers of protons and neutrons in the nucleus.

Due to high value of the nuclear matter incompressibility the total nuclear density $\rho=\rho_n+\rho_p$ in the nuclear volume is constant $\rho=\rho_{0n}+\rho_{0p}$, therefore $\delta\rho_n=- \delta\rho_p$, see also \cite{strutinsky,D89}.

We should take into account that the numbers of protons and neutrons in deformed nucleus are, respectively, $Z$ and $N$; and the center of mass must lie in the plane of mirror symmetry of the nucleus \cite{strutinsky,D89,D92}, because the reflection asymmetric nuclear shapes are coupled by sub-barrier tunnel transition. These two conditions can be easy fulfilled by introduction of auxiliary monopole $\beta_0$ and dipole $\beta_1$ deformations, i.e.
\begin{eqnarray}
\frac{R_p(\theta)}{R_{0p}} = \frac{R_n(\theta)}{R_{0n}} = F(\theta) +\beta_0 Y_{0 0}(\theta) + \beta_1 Y_{1 0}(\theta) \nonumber \\ = f(\theta) = 1+\sum_{\ell=0}^L \beta_\ell Y_{\ell 0}(\theta) .
\end{eqnarray}
The values of $\beta_0$ and $\beta_1$ are, correspondingly, determined by equations
\begin{eqnarray}
\int dV \; \frac{\rho_{0p}}{Z} = \int dV \; \frac{\rho_{0n}}{N}
=\frac{1}{2} \int_0^\pi d\theta\; \sin{(\theta)} f(\theta)^3 = 1 , \\
\int dV \; r \cos{(\theta)} (\rho_{0p} + \rho_{0n}) = \frac{3}{8} (Z R_{0p}+N R_{0n}) \nonumber \\ \times \int_0^\pi d\theta\; \sin{(\theta)} \cos{(\theta)} f(\theta)^4 =0.
\end{eqnarray}

For the sake of simplicity we take into account the most important multipole deformations of nuclear surface $\beta_2$, $\beta_3$, $\beta_4$, $\beta_5$, $\beta_6$. The expressions for $\beta_0$ and $\beta_1$ taken into account all quadratic and cubic terms on $\beta_2$, $\beta_3$, $\beta_4$, $\beta_5$, $\beta_6$ can be directly obtained from Eq. (4) and (5), however corresponding equations are cumbersome and therefore not presented here.

The PEDM is defined as
\begin{eqnarray}
D \equiv e \int dV \; r \cos{(\theta)} \rho_p .
\end{eqnarray}

Due to deviation of the nuclear surface from spherical form there are variation of the proton density into the nuclear volume $\delta \rho_p({\bf r})$. The variation of nucleon density in nuclear volume induces the deviation of the proton radius $\delta R_{p}(\theta)$ from the equilibrium position on the nuclear surface. The proton radius variation induces the proton density variations in the volume $\delta R_{p}(\theta) \Delta {\cal S}$, where $\Delta S$ is the element of surface square. Therefore the PEDM in reflection asymmetric nuclei with axial symmetry is related to the redistribution of protons relatively neutrons into the nuclear volume and on the nuclear surface, see also \cite{DMS,D89},
\begin{eqnarray}
D = D_v + D_s,
\end{eqnarray}
where
\begin{eqnarray}
D_v & \approx & e \int dV r \cos{(\theta)} [\rho_{0p}+\delta \rho_p] = e \int dV r \cos{(\theta)} \delta \rho_p \nonumber \\ & = & 2 \pi e \int_0^\pi d\theta \; \sin{(\theta)} \cos{(\theta)} \int_0^{R_{0p}f(\theta)} dr \; r^3 \delta \rho_p , \\
D_s & \approx & e \int d{\cal S} \; R_{p}(\theta) \cos{(\theta)} \; \rho_{0p} \; \delta R_{p}(\theta) \nonumber \\ & = & \frac{3Ze}{2} \int_0^\pi d\theta \sin{(\theta)} \cos{(\theta)} \left[1+\left(\frac{f^\prime(\theta)}{f(\theta)}\right)^2\right]^{1/2} \nonumber \\ & & \times f^3(\theta) \delta R_{p}(\theta).
\end{eqnarray}
Here Eq.(5) is taken into account at simplification of Eq. (8) and $f^\prime(\theta)=\frac{d f(\theta)}{d\theta}$. So, the volume part of PEDM is related to the volume integral and density variation in the nuclear volume $\delta \rho_p$ while the surface part of the one is determined by the surface integral and the proton radius variation $\delta R_{p}(\theta)$.

The proton (or neutron) density variation induced by surface deformation produces additional pressure on the free nuclear surface. Due to this pressure the position of corresponding surface is slightly shifted. Both the surface symmetry energy and Coulomb force counteract the surface shift and neutralize the additional pressure on the free nuclear surface induced by density variations, see for details \cite{DMS,D89,DD96}. Normal to the surface variation of the proton radius is defined by the boundary condition \cite{DMS,D89,DD96}, which equalizes the normal to surface pressures induced by density fluctuations, neutron-skin stiffness and Coulomb interaction, and equals to
\begin{eqnarray}
\delta R_{p}(\theta) = - \frac{N}{A}
\frac{3 e R_0}{8 Q A^{1/3}} \left[ \phi(R_{p}(\theta)) - \frac{\int d{\cal S} \phi(R_{p}(\theta))}{\int d{\cal S}} \right] \nonumber \\
= - \frac{N}{A}
\frac{3 e R_0}{8 Q A^{1/3}} \left[ \varphi(R_{p}(\theta)) - \frac{\int d{\cal S} \varphi(R_{p}(\theta))}{\int d{\cal S}} \right], \;\;
\end{eqnarray}
where $Q$ is the neutron-skin stiffness coefficient \cite{DMS,DD96}, $\phi=\varphi-\overline{\varphi}$, $\varphi({\bf r})$ is the Coulomb potential related to the protons, $\overline{\varphi}=\frac{\int dV \varphi}{\int dV}$ is the average potential value in the nucleus and $A=Z+N$. Note that $Z \delta R_{p}(\theta) +N \delta R_{n}(\theta) = 0$, because the center of the mass must lie in the plane of mirror symmetry of the nucleus, i.e.
\begin{eqnarray*}
\int dV \; r \cos{(\theta)} \; \rho({\bf r}) \;\;\;\;\;\;\;\;\;\;\;\;\;\;\;\;\;\;\;\;\;\;\;\;\;\;\;\;\;\;\;\;\;\;\;\;\;\;\;\;\;\;\;\;\;\;\;\;\;\; \\ = \int dV \; r \cos{(\theta)} \; [\rho_{0p}+\delta\rho_{p}+\rho_{0n}+\delta\rho_{n}] \;\;\;\;\; \;\;\;\;\; \\ +\int d{\cal S} \; \cos{(\theta)} \; [ R_{p}(\theta) \rho_{0p} \delta R_{p}+R_{n}(\theta) \rho_{0n} \delta R_{n}] \\ = \int dV \; r \cos{(\theta)} \; [\delta\rho_{p}+\delta\rho_{n}] \;\;\;\;\;\;\;\;\;\;\;\;\;\;\;\;\;\;\;\;\;\;\;\;\;\;\; \\ +\int d{\cal S} \; \cos{(\theta)} \; [R_{p}(\theta) \rho_{0p} \delta R_{p}+ R_{n}(\theta) \rho_{0n} \delta R_{n}] \\ = \int d{\cal S} \; \cos{(\theta)} \; [R_{p}(\theta) \rho_{0p} \delta R_{p}+ R_{n}(\theta) \rho_{0n} \delta R_{n}] \\ \approx \int d{\cal S}\; \cos{(\theta)} \; R_{p}(\theta) [\rho_{0p} \delta R_{p}+\rho_{0n} \delta R_{n}]=0.
\end{eqnarray*}
Here Eq. (5) and condition $\delta\rho_{p}=-\delta\rho_{n}$ are taken into account.

If we know $\delta \rho_p({\bf r})$ and $\varphi({\bf r})$ than we can evaluate the PEDM using Eqs. (7)-(10). Let us find $\delta \rho_p({\bf r})$ and $\varphi({\bf r})$ in the framework of liquid-drop model. The energy density functional, which is described density distribution in the nuclear volume, can be written in a simple form \cite{strutinsky,D89,DD96}
\begin{eqnarray}
{\cal E} \approx -a_v \rho + J \frac{(\rho_n-\rho_p)^2}{\rho} + e \rho_p \varphi \nonumber \\ = -a_v \rho + J \frac{(\rho-2\rho_p)^2}{\rho} + e \rho_p \varphi,
\end{eqnarray}
where $-a_v$ is the bulk energy per nucleon in symmetric nuclear matter and $J$ is the volume symmetry energy. Note that the energy density functional of the droplet model contains the dilatation term \cite{DMS}, but parameter $L$ related to the dilatation term is equal zero in recent parameter set of the droplet model \cite{DMS}, therefore we neglect dilatation term here. The energy of nucleus $E$ is related to the energy density functional $E=\int dV {\cal E}$. The equation determined the equilibrium distribution of the charge into the nuclear volume can be obtained by variation of the energy
\begin{eqnarray}
\delta E &=&\delta \int dV \; [{\cal E}-\lambda \rho_p] \\ & = &\int dV \; [ -4 J (\rho-2\rho_p)/\rho + e \varphi - (a_v+\lambda)] \; \delta\rho_p \nonumber
\end{eqnarray}
on $\delta\rho_p$ with the additional condition conserved the number of protons in the nucleus. As the result, we get
\begin{eqnarray}
8J \rho_p =- \rho (e \varphi -4J -\lambda^\prime ) ,
\end{eqnarray}
where $\lambda^\prime=a_v+\lambda$ and $\lambda$ is the Lagrangian coefficient related to the additional condition. The solution of this equation is
\begin{eqnarray}
\rho_{0p} & = & \rho \left(\frac{1}{2} +\frac{\lambda^\prime}{8J} - \frac{e \overline{\varphi}}{8J} \right) \nonumber \\ & = & (\rho_{0p}+\rho_{0n}) \left(\frac{1}{2} +\frac{\lambda^\prime}{8J} - \frac{e \overline{\varphi}}{8J} \right) , \\
\delta \rho_{p} & = & \frac{-e \rho (\varphi-\overline{\varphi})}{8J} = \frac{-e (\rho_{0p}+\rho_{0n}) (\varphi-\overline{\varphi})}{8J}\nonumber \\ & = & \frac{-3e A (\varphi-\overline{\varphi})}{32 \pi R_{0p}^3 J} = \frac{-3e A \phi}{32 \pi R_{0p}^3 J}.
\end{eqnarray}
Note that $\int dV \delta \rho_{p} = \frac{-3e A}{32 \pi R_{0p}^3 J} \left[ \int dV (\varphi-\overline{\varphi})\right]=0 $.
We neglect by the difference between $R_{0p}$ and $R_{0n}$ at evaluation of volume quantities.

The electric potential can be found by using the Poisson equation
\begin{eqnarray}
\nabla^2\varphi = 4 \pi e \rho_p.
\end{eqnarray}

Substituting (15) into (8) and taking into account (5) we get
\begin{eqnarray}
D_v \approx \frac{-3 e^2 A}{16 J R_{0p}^3} \int_0^\pi d\theta \; \sin{(\theta)} \cos{(\theta)} \int_0^{R_{0p}f(\theta)} dr \; r^3 \phi({\bf r}) \nonumber \\
= \frac{-3 e^2 A}{16 J R_{0p}^3} \int_0^\pi d\theta \; \sin{(\theta)} \cos{(\theta)} \int_0^{R_{0p}f(\theta)} dr \; r^3 \varphi({\bf r}) . \;\;\;\;
\end{eqnarray}
Using (10) and approximation $\frac{NZ}{A}\approx \frac{A}{4}$, see Ref. \cite{DMS}, we rewrite (9) into the form
\begin{eqnarray}
D_s & \approx & -\frac{9Ae^2 R_0}{64 Q A^{1/3}} \int_0^\pi d\theta \; g(\theta) \cos{(\theta)} f(\theta) \\ & & \times
\left[ \varphi(R_{p}(\theta)) - \frac{\int_0^\pi d\theta^\prime \; g(\theta^\prime) \varphi(R_{p}(\theta^\prime))}{\int_0^\pi d\theta^\prime \; g(\theta^\prime)} \right] \nonumber,
\end{eqnarray}
where $g(\theta)=\sin{(\theta)} f^2(\theta) \left[1+\left(f^\prime(\theta)/f(\theta)\right)^2\right]^{1/2}$.
So, the volume (17) and surface (18) parts of PEDM are determined by the Coulomb potential $\varphi({\bf r})$.

The Coulomb potential of deformed nucleus is
\begin{eqnarray}
\varphi({\bf r}) = e \int dV \frac{\rho_{p}({\bf r}^\prime)}{|{\bf r} - {\bf r}^\prime|} = e \int dV \frac{\rho_{0p}+\delta \rho_{p}({\bf r}^\prime)}{|{\bf r} - {\bf r}^\prime|} .
\end{eqnarray}
This potential satisfies to Eq. (16).

It is possible to find potential $\varphi({\bf r})$ by applying the perturbation theory to Eqs. (14), (15) and (19). We expand the potential and the variation of proton density into the perturbation series
\begin{eqnarray}
\varphi ({\bf r})= \varphi^{0}({\bf r}) +\varphi^{1}({\bf r}) +\varphi^{2}({\bf r}) +... \; , \\
\delta \rho_{p}({\bf r})=\delta \rho_{p}^{0}({\bf r}) +\delta \rho_{p}^{1}({\bf r}) + \delta \rho_{p}^{2}({\bf r}) +... \;,
\end{eqnarray}
where the superscript corresponds to the number of perturbation approach. Corresponding solution for the Lagrangian coefficient is variation of proton density into the perturbation series
\begin{eqnarray*}
\lambda=-a_v+8J[Z/A-1/2+ e/(8J)(\overline{\varphi^{0}({\bf r})} \\ + \overline{\varphi^{1}({\bf r})} + \overline{\varphi^{2}({\bf r})} +... )].
\end{eqnarray*}

Substituting perturbation series (20)-(21) into Eqs. (15) and (19) we get
\begin{eqnarray}
\delta \rho_{p}^k({\bf r}) = \frac{-e \rho \phi^k({\bf r})}{8J} = \frac{-3e A \phi^k({\bf r})}{32 \pi R_{0p}^3 J} , \;\; {\rm for} \; k \geq 0 ,
\;\; \\
\varphi^k({\bf r}) = e \int dV \frac{\delta \rho_p^{k-1}({\bf r}^\prime)}{|{\bf r} - {\bf r}^\prime|}, \;\; {\rm for} \; k \geq 1 , \;\;
\end{eqnarray}
and
\begin{eqnarray}
\varphi^{0}({\bf r})= e \int dV \frac{\rho_{0p} }{|{\bf r} - {\bf r}^\prime|} = \frac{eZ}{R_{0p} }\int dV \frac{3}{4 \pi R_{0p}^2|{\bf r} - {\bf r}^\prime|} \nonumber \\
= \frac{eZ}{R_{0p}} \sum_\ell \frac{6 \pi Y_{\ell0}(\theta)}{(2\ell+1)} \int_0^\pi d\theta^\prime \sin(\theta^\prime) Y_{\ell0}^*(\theta^\prime) \;\;\; \nonumber \\ \times \left[ \int_0^r dr^\prime \frac{(r^\prime)^{\ell+2} }{r^{\ell+1}(R_{0p})^2} + \int_r^{R_{0p}f(\theta^\prime) } dr^\prime \frac{(r)^{\ell} }{(r^\prime)^{\ell-1}(R_{0p})^2}\right], \;\;
\end{eqnarray}
where $\phi^k({\bf r})=\varphi^k({\bf r})-\overline{\varphi^k({\bf r})}$.

Using Eqs. (22) and (23) we get the recurrent equation for $\varphi^k({\bf r})$ at $ k \geq 1 $
\begin{eqnarray}
\varphi^k({\bf r}) & = & \frac{-3e^2 A}{32 \pi R_{0p}^3 J} \int dV \frac{ \phi^{k-1}({\bf r}^\prime)}{|{\bf r} - {\bf r}^\prime|} \nonumber \\
& = & \frac{-e^2 A}{R_{0p} J} \sum_\ell \frac{3 \pi Y_{\ell0}(\theta)}{4(2\ell+1)} \int_0^\pi d\theta^\prime \sin(\theta^\prime) Y_{\ell0}^*(\theta^\prime) \;\;\;\;\; \nonumber \\ & &
\times \left[ \int_0^r dr^\prime \frac{(r^\prime)^{\ell+2} \phi^{k-1}({\bf r}^\prime)}{r^{\ell+1}(R_{0p})^2} \right. \nonumber \\ & & \left. \;\;\;\;\;
+ \int_r^{R_{0p}f(\theta^\prime) } dr^\prime \frac{(r)^{\ell} \phi^{k-1}({\bf r}^\prime)}{(r^\prime)^{\ell-1}(R_{0p})^2}\right],
\end{eqnarray}
which determines the potential with any necessary degree of accuracy. As a result, we can evaluate the volume and surface contributions of PEDM using Eqs. (17), (18), (20), (24) and (25).

The macroscopic PEDM can be written as
\begin{eqnarray}
D_{\rm macro} = D_{v 1} + D_{v 2 0} +D_{v 2 1} + D_{s 1} + D_{s 2 0} +D_{s 2 1}, \;\;
\end{eqnarray}
where
\begin{eqnarray}
D_{v 1}=\frac{e^3 AZ}{\pi J} \left[\frac{9 \beta_2 \beta_3 }{56 \sqrt{35}}+\frac{11 \beta_3 \beta_4 }{105 \sqrt{7} } \right. \nonumber \\ \left. + \frac{41 \beta_4 \beta_5 }{264 \sqrt{11}} +\frac{441 \beta_5 \beta_6 }{715 \sqrt{143} } \right], \\
D_{s 1}= \frac{15e^3 A^{2/3} Z}{8 \pi Q} \left[\frac{9 \beta_2 \beta_3 }{56 \sqrt{35}}+\frac{11 \beta_3 \beta_4 }{105 \sqrt{7} } \right. \nonumber \\ \left. +\frac{41 \beta_4 \beta_5 }{264 \sqrt{11}} +\frac{441 \beta_5 \beta_6 }{715 \sqrt{143} } \right]
\end{eqnarray}
are the volume and surface first-order contributions,
\begin{eqnarray}
D_{v 2 0}= \frac{e^3 AZ}{\pi^{3/2} J} \left[
\frac{3 \beta_2^2 \beta_3 }{56 \sqrt{7} }
+\frac{789 \beta_2^2 \beta_5 }{8624 \sqrt{11} }
+\frac{48721 \beta_2 \beta_3 \beta_4 }{101640 \sqrt{35} } \right. \nonumber \\
+\frac{65685 \beta_2 \beta_3 \beta_6 }{44044 \sqrt{455} }
+\frac{1658135 \beta_2 \beta_4 \beta_5 }{2186184 \sqrt{55}}
+\frac{35403 \beta_2 \beta_5 \beta_6 }{11440 \sqrt{715}} \nonumber \\
+\frac{3 \beta_3^3 }{88 \sqrt{7} }
+\frac{19557 \beta_3^2 \beta_5 }{104104 \sqrt{11} }
+\frac{27147 \beta_3 \beta_4^2 }{220220 \sqrt{7} }
+\frac{657095 \beta_3 \beta_4 \beta_6 }{528528 \sqrt{91} } \nonumber \\
+\frac{141723 \beta_3 \beta_5^2 }{1041040 \sqrt{7} }
+\frac{110793 \sqrt{7} \beta_3 \beta_6^2 }{5348200 }
+\frac{245625 \beta_4^2 \beta_5 }{1457456 \sqrt{11} } \nonumber \\ \left.
+\frac{46892 \beta_4 \beta_5 \beta_6 }{36465 \sqrt{143} }
+\frac{327 \beta_5^3 }{5746 \sqrt{11} }
+\frac{64461 \beta_5 \beta_6^2 }{369512 \sqrt{11} } \right], \;\;
\end{eqnarray}
\begin{eqnarray}
D_{s 2 0}=\frac{e^3 A^{2/3} Z}{\pi^{3/2} Q} \left[
\frac{297 \beta_2^2 \beta_3 }{2240 \sqrt{7} }
+\frac{20277 \beta_2^2 \beta_5 }{68992 \sqrt{11} } \right. \nonumber \\
+\frac{80181 \beta_2 \beta_3 \beta_4 }{54208 \sqrt{35} }
+\frac{2047545 \beta_2 \beta_3 \beta_6 }{352352 \sqrt{455}}
+\frac{16455195 \beta_2 \beta_4 \beta_5 }{5829824 \sqrt{55} } \nonumber \\
+\frac{252207 \beta_2 \beta_5 \beta_6 }{18304 \sqrt{715} }
+\frac{81 \beta_3^3 }{704 \sqrt{7} }
+\frac{56025 \beta_3^2 \beta_5 }{75712 \sqrt{11} }
+\frac{177669 \beta_3 \beta_4^2 }{352352 \sqrt{7} } \nonumber \\
+\frac{8432595 \beta_3 \beta_4 \beta_6 }{1409408 \sqrt{91} }
+\frac{1113129 \beta_3 \beta_5^2 }{1665664 \sqrt{7} }
+\frac{1037259 \sqrt{7} \beta_3 \beta_6^2 }{8557120 } \nonumber \\
+\frac{9802305 \beta_4^2 \beta_5 }{11659648 \sqrt{11} }
+\frac{299061 \beta_4 \beta_5 \beta_6 }{38896 \sqrt{143} } \nonumber \\ \left.
+\frac{31455 \beta_5^3 }{91936 \sqrt{11} }
+\frac{3679965 \beta_5 \beta_6^2 }{2956096 \sqrt{11} } \right] \;\;\;
\end{eqnarray}
are the volume and surface second-order contributions related to $\varphi^0({\bf r})$ contribution (see Eqs. (17), (18), (20), (24)), and
\begin{eqnarray}
D_{v 2 1}= - \frac{e^5 A^{5/3} Z}{\pi J^2 r_0} \left[
\frac{477 \beta_2 \beta_3 }{15680 \sqrt{35} }
+\frac{3719 \beta_3 \beta_4 }{194040 \sqrt{7} } \right. \nonumber \\ \left.
+\frac{176933 \beta_4 \beta_5 }{6342336 \sqrt{11} }
+\frac{627219 \beta_5 \beta_6 }{5725720 \sqrt{143} } \right] , \\
D_{s 2 1}=- \frac{e^5 A^{4/3} Z}{\pi J Q r_0} \left[
\frac{459 \beta_2 \beta_3 }{7840 \sqrt{35} }
+\frac{9623 \beta_3 \beta_4 }{258720 \sqrt{7} } \right. \nonumber \\ \left.
+\frac{32881 \beta_4 \beta_5 }{604032 \sqrt{11} }
+\frac{702081 \beta_5 \beta_6 }{3271840 \sqrt{143}} \right]
\end{eqnarray}
are the volume and surface second-order contributions connected to $\varphi^1({\bf r})$ contribution (see Eqs. (17), (18), (19), (25)), $r_0=R_{0p}/A^{1/3}$. This expression of the PEDM is obtained with the help of symbolic computation software $Mathematica$.

We propose at evaluation of the PEDM that ratio of potentials $\varphi^1({\bf r})/\varphi^0({\bf r})$ is the same order as $\beta_\ell$, therefore we take into account terms proportional to the product of deformations $\beta_\ell \beta_{\ell^\prime}$ in $D_{v 2 1}$ or $D_{s 2 1}$ and neglect by the next order terms $\beta_\ell \beta_{\ell^\prime} \beta_{\ell^{\prime \prime}}$.
This proposal is natural for the hierarchy of solutions in the form of the perturbation series.

The first term in Eq. (27) was obtained in Refs. \cite{strutinsky,b-m2}, Eqs. (27) and (28) were derived in Refs. \cite{DMS,D89}, and Eqs. (29)-(32) are found for the first time.

\section{Discussion and conclusions}

\begin{figure}
\resizebox{0.5\textwidth}{!}{%
\includegraphics{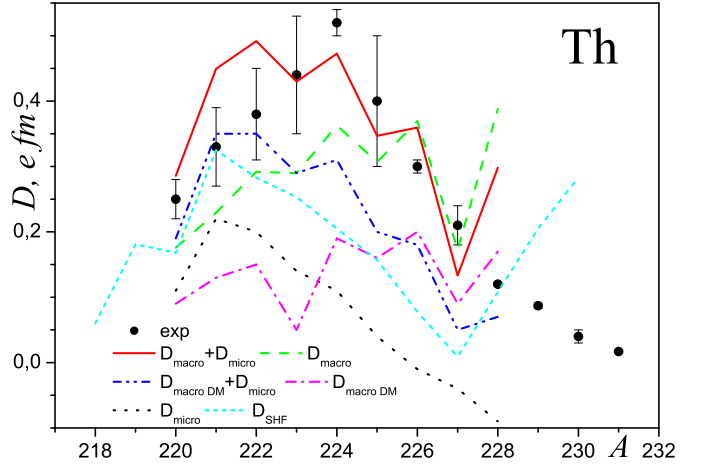}
}
\caption{Experimental and theoretical values of the PEDM as well as macroscopic and microscopic contributions into the PEDM for Th isotopes. Points are experimental data from Refs. \cite{BN-RMP,th227-231a,th227-231b}. Details for theoretical lines see in text.}
\label{fig:1}    
\end{figure}

The total value of PEDM $D_{\rm tot}$ is the sum of macroscopic $D_{\rm macro}$ and microscopic shell-correction $D_{\rm micro}$ contributions \cite{D92,BN-D,Skalski} calculated for the same shapes of the proton and neutron surfaces \cite{D92}, i.e.
\begin{eqnarray}
D_{\rm tot} = D_{\rm macro}+D_{\rm micro}.
\end{eqnarray}
The macroscopic part of PEDM can be evaluated by using Eqs. (26)-(32).

\begin{figure}
\resizebox{0.5\textwidth}{!}{%
\includegraphics{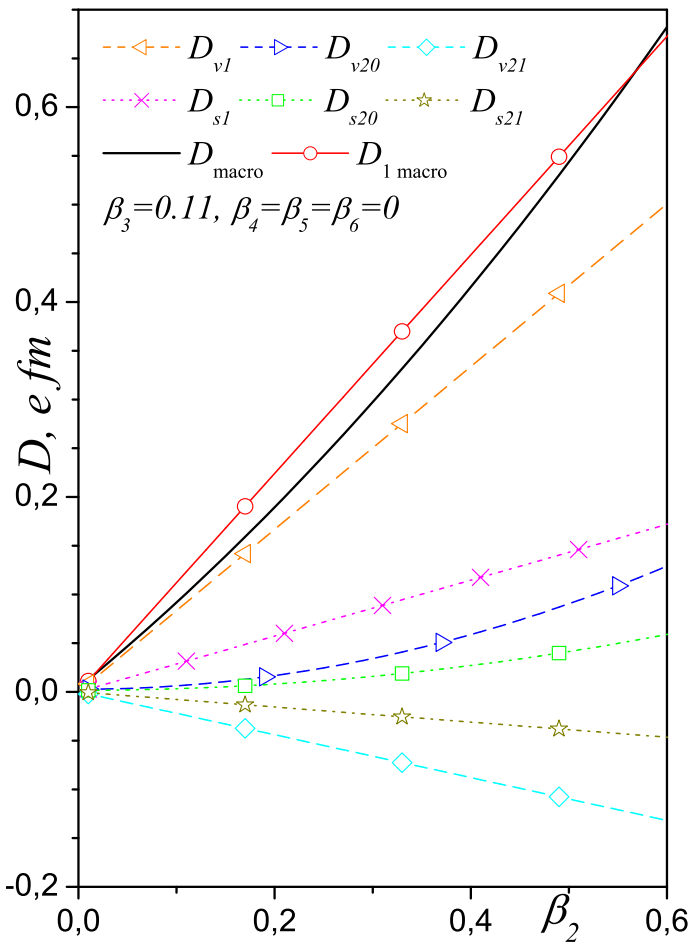}
}
\caption{Dependencies of the total macroscopic PEDM evaluated in the first $D_{\rm 1 macro}$ and second $D_{\rm macro}$ orders on quadrupole $\beta_2$ deformation as well as the same dependencies of contributions $D_{v 1}$, $D_{v 2 0}$, $D_{v 2 1}$, $D_{s 1}$, $D_{s 2 0}$, and $D_{s 2 1}$ into the PEDM. The quadrupole and octupole deformations of $^{220}$Th are only taken into account.}
\label{fig:2}    
\end{figure}

\begin{figure}
\resizebox{0.5\textwidth}{!}{%
\includegraphics{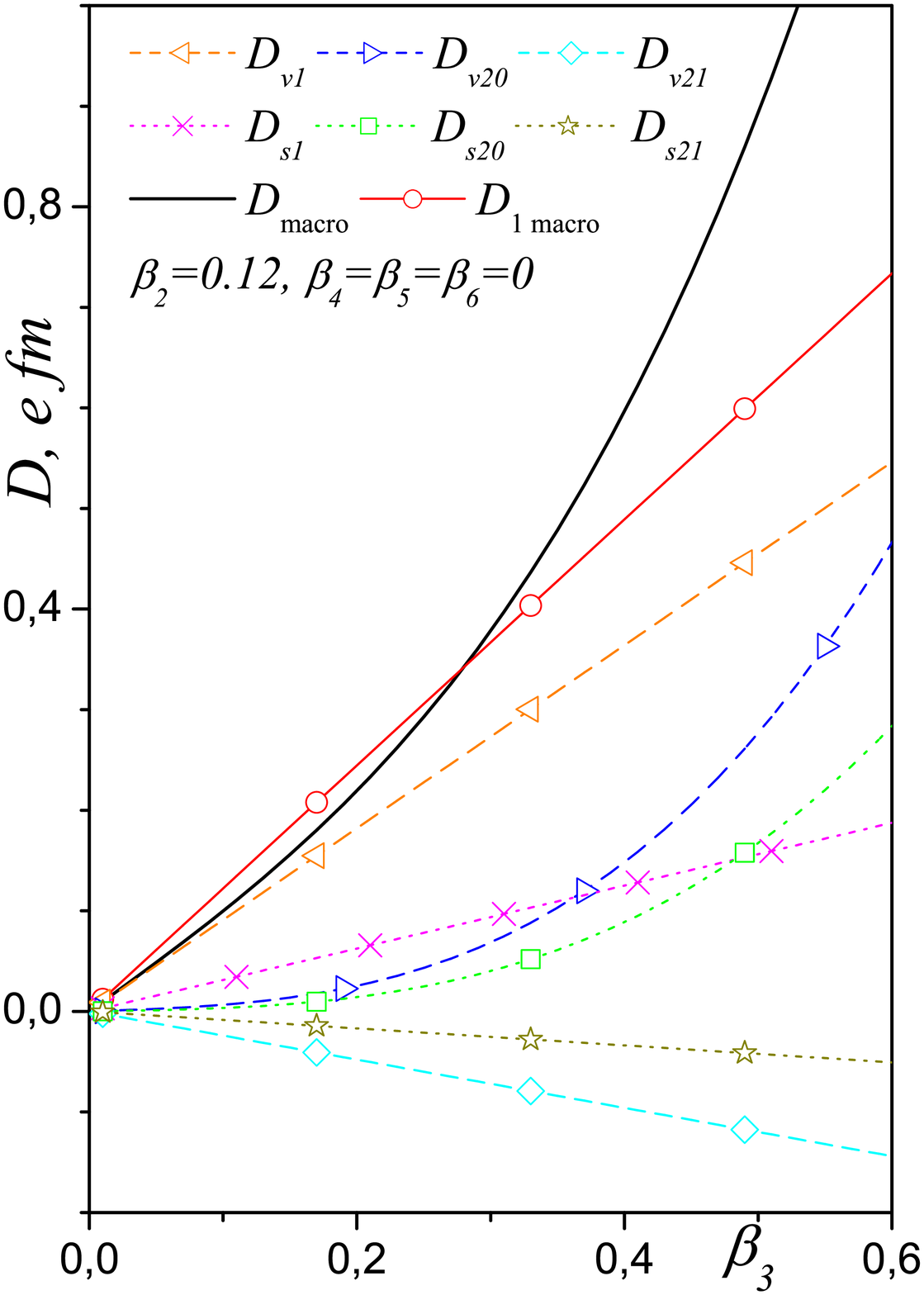}
}
\caption{Dependencies of the total macroscopic PEDM evaluated in the first $D_{\rm 1 macro}$ and second $D_{\rm macro}$ orders on octupole $\beta_3$ deformation as well as the same dependencies of contributions $D_{v 1}$, $D_{v 2 0}$, $D_{v 2 1}$, $D_{s 1}$, $D_{s 2 0}$, and $D_{s 2 1}$ into the PEDM. The quadrupole and octupole deformations of $^{220}$Th are only taken into account.}
\label{fig:3}    
\end{figure}

\begin{figure}
\resizebox{0.5\textwidth}{!}{%
\includegraphics{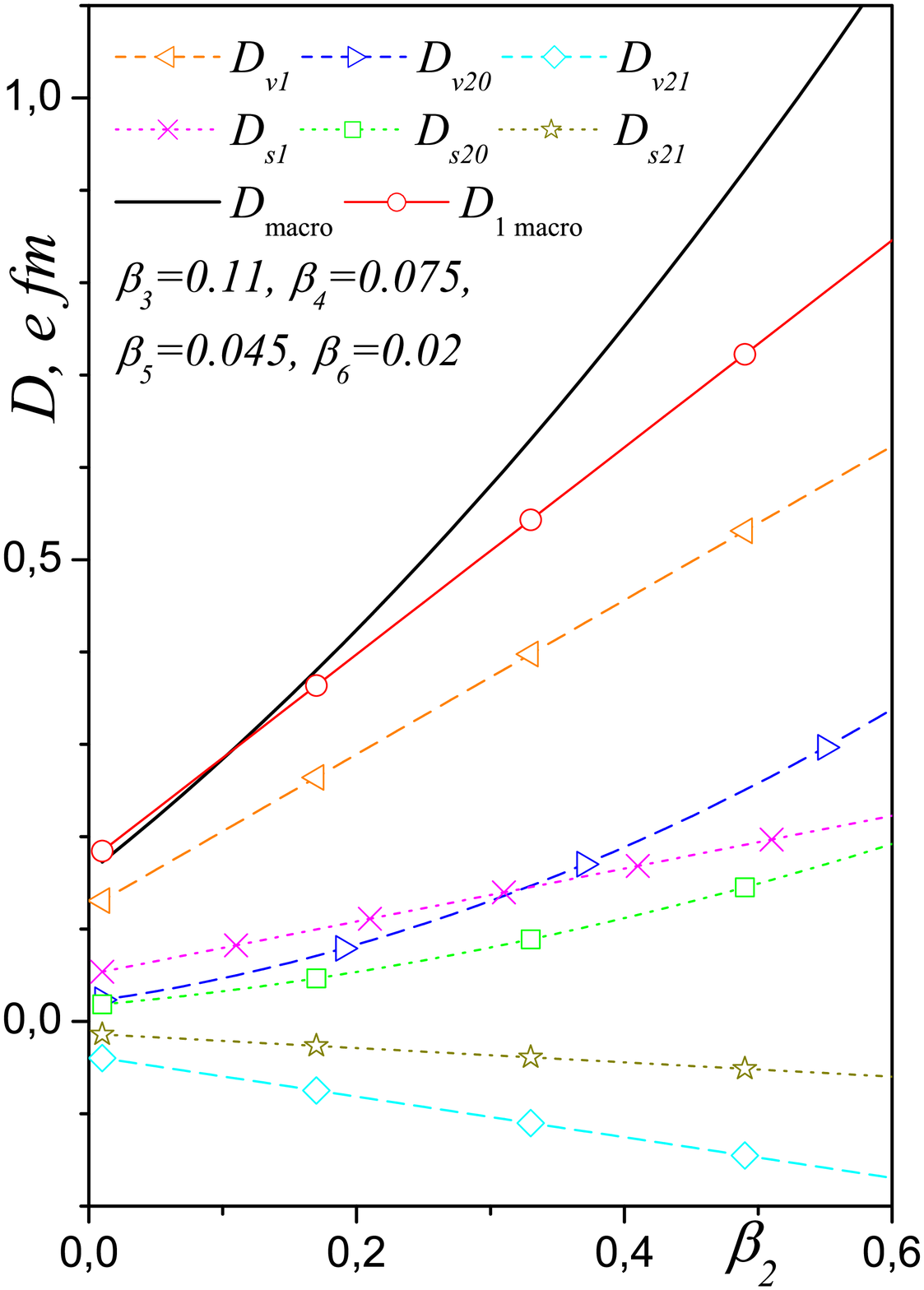}
}
\caption{The same as in Fig. 2, but the high-order multipole deformations $\beta_4$, $\beta_5$ and $\beta_6$ of $^{220}$Th are also taken into account.}
\label{fig:4}    
\end{figure}

\begin{figure}
\resizebox{0.5\textwidth}{!}{%
\includegraphics{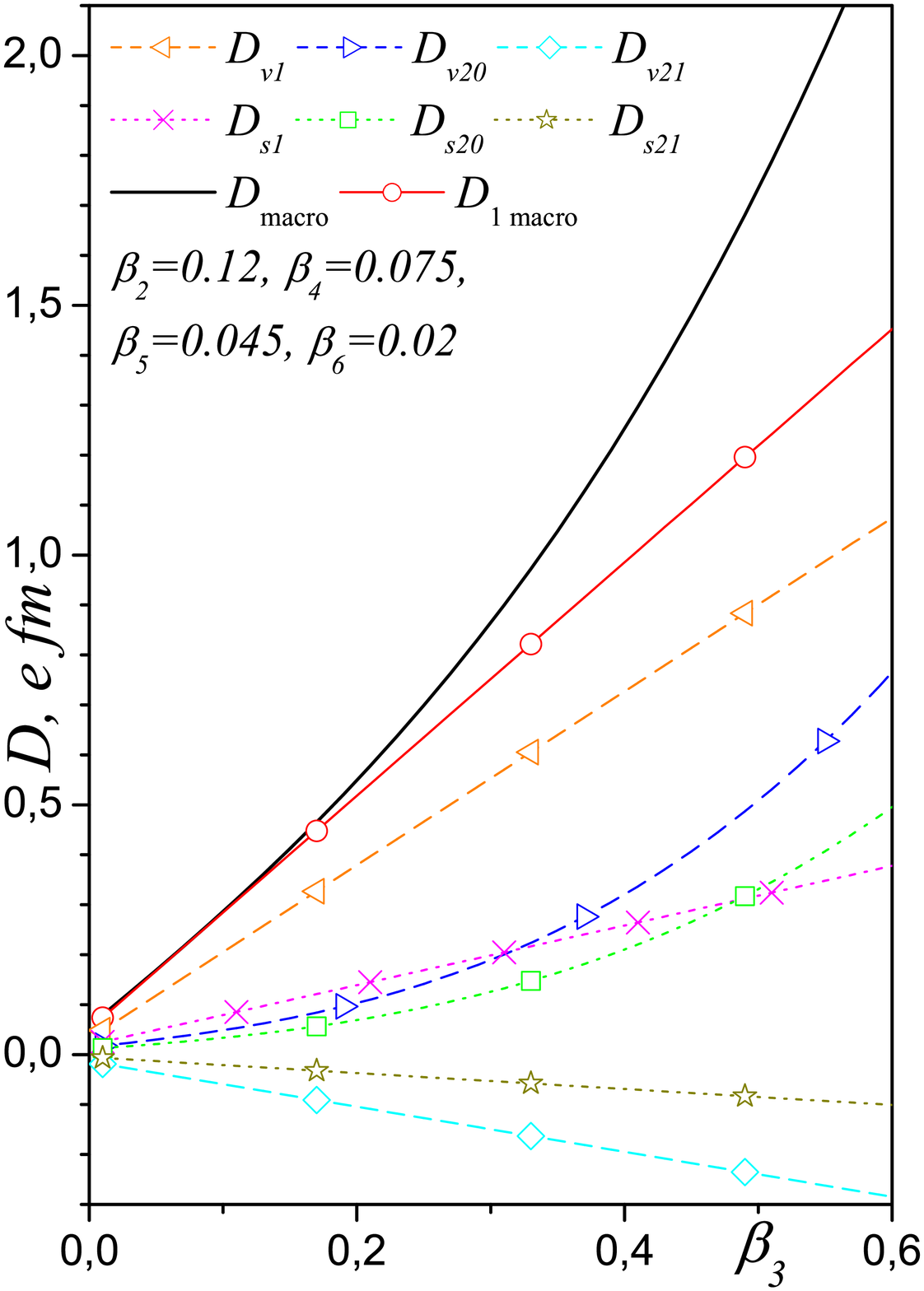}
}
\caption{The same as in Fig. 3, but the high-order multipole deformations $\beta_4$, $\beta_5$ and $\beta_6$ of $^{220}$Th are also taken into account.}
\label{fig:5}    
\end{figure}

The total values of PEDM evaluated in the framework of various models are compared with the experimental data for thorium isotopes in Fig. 1. The experimental data are taken from Refs. \cite{BN-RMP,th227-231a,th227-231b}. Our calculation of the macroscopic part $D_{\rm macro}$ is done with the help of Eqs. (26)-(32) using recent parameter values of the droplet model $J=32.5$ MeV, $Q=29.4$ MeV, $r_0=1.16$ \cite{DMS}. The values of multipole deformation parameters $\beta_\ell$ and microscopic part of PEDM $D_{\rm micro}$ are taken from Ref. \cite{BN-D}. The results obtained in our model well agree with the experimental data for $^{220-228}$Th, see Fig. 1. The total values of PEDM calculated by Butler and Nazarewicz using the droplet model approach for macroscopic part $D_{\rm tot \; BN}=D_{\rm macro \; DM}+D_{\rm micro}$ \cite{BN-D} are shown in Fig. 1 too. The results of the droplet model approach underestimate the experimental data for $^{223-228}$Th, see Fig. 1. The values of PEDM obtained in the framework of cranking Skyrme-Hartree-Fock approach $D_{\rm SHF}$ \cite{TsKN} are also presented in Fig. 1. The values of PEDM evaluated in the cranking Skyrme-Hartree-Fock model underestimate the experimental data for $^{222-227}$Th and overestimate the ones for $^{229,230}$Th. Comparison of the PEDM values calculated in the framework of various models with the experimental data for thorium isotopes in Fig. 1 suggest that our proposal, that the proton and neutron surfaces are geometrically similar (see Eq. (2)), is reasonable.

The values of macroscopic part of PEDM evaluated in our model $D_{\rm macro}$ are larger than the ones obtained in the framework of the droplet model $D_{\rm macro \; DM}$, see Fig. 1 and Refs. \cite{DMS,MS,D89,D92}. Comparing the results in Fig. 1 we conclude that the microscopic contribution to the PEDM $ D_{\rm micro}$ is smaller than the macroscopic one $D_{\rm macro}$.

Let us study the role of second-order contributions into the macroscopic part of the PEDM in well-deformed nuclei. We consider nuclei with quadrupole and octupole deformations at the beginning. Here we neglect by the microscopic part of PEDM for simplicity. Dependence of the PEDM on the quadrupole deformation value at fixed value of octupole deformation is presented in Fig. 2, while dependence of the PEDM on the octupole deformation value at fixed value of quadrupole deformation is presented in Fig. 3. Fixed values of quadrupole or octupole deformations are pointed in Figs. 2-3. These values of the deformation parameters are typical for the ground-state of reflection asymmetric actinides \cite{BN-D}.

The macroscopic PEDM consists of six contributions $D_{v 1}$, $D_{v 2 0}$, $D_{v 2 1}$, $D_{s 1}$, $D_{s 2 0}$, and $D_{s 2 1}$, see Eqs. (26)-(32). Contributions of all these terms into the PEDM as well as the total first and second orders macroscopic PEDM values are shown in Figs. 2-3. Comparing various lines in Figs. 2-3 we conclude.
\begin{itemize}

\item[-] The total first order contribution of the PEDM $D_{\rm 1 macro}=D_{v 1} + D_{s 1}$ is mainly determined the value of PEDM at small values of the deformation parameters. The influence of the second order terms rises with the values of the deformation parameters.

\item[-] Contributions $D_{v 1}$, $D_{s 1}$, $D_{\rm 1 macro}$, $D_{v 2 1}$, and $D_{s 2 1}$ of the PEDM depend on variable deformation parameter linearly, while $D_{\rm macro}$, $D_{v 2 0}$ and $D_{s 2 0}$ depend on variable deformation parameter quadratically.

\item[-] Surface contribution of any type is approximately twice smaller than the volume contribution of the same type. (Note that this conclusion depends on the ratio between values $J$ and $Q$. We would remind that we use the same values of $J$ and $Q$ as in the droplet model at PEDM evaluation, but some authors \cite{DMS,BN-D,Skalski} use other values $J$ and $Q$ at description of the PEDM experimental data.)

\item[-] Terms $D_{v 2 1}$ and $D_{s 2 1}$ related to the Coulomb potential correction $\varphi^1$ give negative contributions into the PEDM, while any other contributions are positive.

\item[-] The absolute values of terms $D_{v 2 1}$ and $D_{s 2 1}$ are similar to the ones for terms $D_{v 2 0}$ and $D_{s 2 0}$. This is support our proposal on the hierarchy of solutions of the perturbation series.

\item[-] The contribution of second order terms $D_{v 2 0} +D_{v 2 1} + D_{s 2 0} +D_{s 2 1}$ gives small correction for the case of variable quadrupole and fixed octupole deformations (see Fig. 2). However this contribution noticeably enhances the value of PEDM at large octupole deformation in the case of variable octupole and fixed quadrupole deformations (see Fig. 3). The total PEDM evaluated at large octupole and fixed quadrupole deformations is large than the one for large quadrupole and fixed octupole deformations.

\end{itemize}

Nuclei with reflection asymmetry have also non-zero values of high-order multipole deformations \cite{Skalski,BN-D}. Dependencies of macroscopic PEDM on the value quadrupole and octupole deformation at fixed values of other deformations are presented in Figs. 4 and 5 respectively. Values of fixed high-order multipole deformations parameters pointed in Figs. 4-5 are typical for the ground-state of reflection asymmetric actinides \cite{BN-D}.

The values of high-order multipole deformations are smaller than the ones for quadrupole or octupole deformations as a rule \cite{Skalski,BN-D}. Nevertheless high-order multipole deformations enhance the value of PEDM noticeably, compare results presented in Figs. 2-5. The influence of second order terms is also strengthening by high-order multipole deformations.

Qualitatively similar results related to the influence of second order contribution on the PEDM was obtained numerically in Ref. \cite{Skalski}, but contributions $D_{v 2 1}$ and $D_{s 2 1}$ related to $\varphi^1$ were skipped in this work. Note that Skalski \cite{Skalski} evaluate the macroscopic part of the PEDM for the case of constant neutron skin (the droplet model approach). As pointed in the introduction the PEDM evaluated in the framework of droplet model has additional contribution related to the difference between the center of mass of neutron skin of uniform thickness and the center of mass of the nucleus, which reduces the value of the PEDM induced by density redistribution \cite{DMS}, compare also lines denoted as $D_{\rm macro}$ and $D_{\rm macro \; DM}$ in Fig. 1. This neutron skin contribution into the PEDM is zero in the case of neutron skin formed by geometrically similar neutron and proton surfaces. Therefore numerical comparison of our and Skalski's results cannot show the difference between the first and second order contributions into the PEDM obtained in various approaches, because the comparison shows mainly the difference related to the neutron-skin contribution.

We evaluate the PEDM of the hyperdeformed state of $^{152}$Dy. The values of deformation parameters of $^{152}$Dy in hyperdeformed state are $\beta_2=0.61$, $\beta_3=0.1$, $\beta_4=0.11$ $\beta_5=0.05$ and $\beta_6=0$ \cite{Skalski}. The values of macroscopic part of PEDM obtained in the first and second orders using Eqs. (26)-(32) are $D_{\rm macro}=0.67 \; e$ fm and $D_{\rm 1 macro}=0.46 \; e$ fm correspondingly. The microscopic shell-correction part of PEDM evaluated for geometrically similar proton and neutron surfaces is $D_{\rm shell}=-0.34$ $e$ fm \cite{Skalski}. As the result the total values of PEDM found in the cases of applying the first and second orders calculation of the macroscopic part of PEDM are  $D_{\rm macro}+D_{\rm shell}=0.33 \; e$ fm and $D_{\rm 1 macro}+D_{\rm shell}=0.12 \; e$ fm respectively. Note that the total values of PEDM evaluated using the exact numerical calculation of the macroscopic contribution in the framework of the droplet model is $D_{\rm macro\; Skalski}+D_{\rm shell}=0.06 \; e$ fm \cite{Skalski}. Hereby, the PEDM depends strongly on the second-order terms in well-deformed nuclei as well as on the neutron skin shape.

In conclusion, the expression for macroscopic PEDM taken into account the first and second order terms on the parameters of multipole deformations is obtained in the case of geometrically similar proton and neutron surfaces of reflection asymmetric nuclei.

The second order terms are important at large values of deformation parameters. The second order terms are especially important for nuclear shape with non-zero values of high-order multipole deformation parameters.

Second order contributions $D_{v 2 0}$ and $D_{s 2 0}$ enlarge the value of PEDM obtained in the first order. In contrast to this contributions $D_{v 2 1}$ and $D_{s 2 1}$ decrease the value of PEDM obtained in the first order. Compensation of these contributions into the PEDM value occurs at small values of deformations. However contribution $D_{v 2 0}+D_{s 2 0}$ into the PEDM value is larger than contributions $D_{v 2 1}+D_{s 2 1}$ at large values of deformation parameters.

The reduced probabilities of dipole transitions $B(E1)$ are measured in various experiments \cite{BN-RMP,expDipMom1a,expDipMom1b,expDipMom2,th227-231a,th227-231b}. Note that $B(E1)$ is proportional to the squared value of the PEDM $D^2$ \cite{BN-RMP,DDa,DDb,DDc,Minkov,RRF}. Therefore the second order contribution into PEDM leads to significant variation of $B(E1)$ in well-deformed nuclei.

The obtained expression can be easily applied for estimation of the macroscopic PEDM and strength of dipole transition probabilities in various well-deformed nuclei.

\section*{Acknowledgements}

Author thanks Prof. F. F. Karpeshin for stimulating discussions.


\begin{thebibliography}{}

\bibitem{strutinsky}
V. M. Strutinsky, At. Energ. {\bf 4}, 150 (1956) (in Russian); J. Nucl. Energy 4, 523 (1957) (English translation).

\bibitem{b-m1} A. Bohr, and B. R. Mottelson, Nucl. Phys. 4, 529 (1957).

\bibitem{b-m2} A. Bohr, and B. R. Mottelson, Nucl. Phys. 9, 687 (1958).

\bibitem{lipas} P. O. Lipas, Nucl. Phys. 40, 629 (1963).

\bibitem{leper} D. P. Leper, Izv. AN SSSR. Ser. Fiz 29, 1253 (1965).

\bibitem{DMS} C. O. Dorso, W. Myers, and W. Swiatecki, Nucl. Phys.
A451, 189 (1986).

\bibitem{MS} W. Myers, and W. Swiatecki, Nucl. Phys.
A531, 93 (1991).

\bibitem{D89} V. Yu. Denisov, Yad. Fiz. 49, 644 (1989) [Sov. J. Nucl. Phys. 49, 399 (1989)].

\bibitem{D92} V. Yu. Denisov, Yad. Fiz. 55, 2647 (1992) [Sov. J. Nucl. Phys. 55, 1478 (1992)].

\bibitem{DD96} V. Yu. Denisov, O. I. Davidovskaya, Yad. Fiz. 59, 981 (1996) [Phys. At. Nucl. 59, 981 (1996)].

\bibitem{Skalski91} J. Skalski, Phys. Rev. C43, 140 (1991).

\bibitem{Skalski} J. Skalski, Phys. Rev. C49, 2011 (1994).

\bibitem{BN-RMP} P. A. Butler and W. Nazarewicz, Rev. Mod. Phys., 68, 350 (1996).

\bibitem{BN-D} P. A. Butler and W. Nazarewicz, Nucl. Phys. A 533, 249 (1991).

\bibitem{TsKN} A. Tsvetkov, J. Kvasil and R. G. Nazmitdinov, J. Phys. G28, 2187 (2002).

\bibitem{BMSV} L. M. Robledo, M. Baldo, P. Schuck, and X. Vinas, Phys. Rev. C 81, 034315 (2010)

\bibitem{DDa} A. Ya. Dzyublik and V. Yu. Denisov, Yad. Fiz. 56, 30 (1993) [Phys. At. Nucl. 56, 303 (1993)]

\bibitem{DDb}V. Yu. Denisov and A. Ya. Dzyublik, Yad. Fiz. 56, 96 (1993) [Phys. At. Nucl. 56, 477 (1993)]

\bibitem{DDc}V. Yu. Denisov and A. Ya. Dzyublik, Nucl. Phys. A 589, 17 (1995).

\bibitem{Minkov} N. Minkov, et al., Phys. Rev. C 76, 034324 (2007).

\bibitem{RRF} A. A. Raduta, C. M. Raduta, and A. Faessler, Phys. Rev. C80, 044327 (2009).

\bibitem{expDipMom1a} T. Rzaca-Urban, et al., Phys. Rev. C 82, 017301 (2010).

\bibitem{expDipMom1b} S. H. Liu, et al., Phys. Rev. C 81, 057304 (2010).

\bibitem{expDipMom2} S. S. Ntshangase, et al., Phys. Rev C 82, 041305(R) (2010).

\bibitem{th227-231a} N. J. Hammond, et al., Phys. Rev. C 65, 064315 (2002). 

\bibitem{th227-231b} K. Gulda, et al., Nucl. Phys. A 703, 45 (2002).

\bibitem{Karpeshin92} F. F. Karpeshin, Z. Phys. 344, 55 (1992).

\bibitem{Karpeshin10} F. F. Karpeshin, Eur. Phys. J. A 45, 251 (2010)

\bibitem{Karpeshin10a}Proc. 3rd Int. Conf. on Nuclear Physics and Atomic Energies, June 2010, Kyiv (Institute for Nuclear Research, Kyiv, 2011), p. 441.

\bibitem{Karpeshin08} F. F. Karpeshin, Proc. 41-42 Winter Schools of St.-Petersburg's Institute of Nuclear Physics, (Institute of Nuclear Physics, St.-Petersburg, 2008), p. 216.

\bibitem{SD-oct1} B. Crowell, et al., Phys. Rev. C51, R1599 (1995).
 
\bibitem{SD-oct2} A. N. Wilson, et al., Phys. Rev. C54, 559 (1996). 

\bibitem{SD-oct3} T. Nakatsukasa, et al., Phys. Rev. C53, 2213 (1996). 

\bibitem{SD-oct4} A. Korichi, et al., Phys. Rev. Lett. 86, 2746 (2001). 

\bibitem{SD-oct5} J. Kvasil, et al., Phys. Rev. C75, 034306 (2007).

\bibitem{KS} M. Kowal and J. Skalski, Phys. Rev. C82, 054303 (2010).

\bibitem{thskin1} A. Bohr and B. Mottelson, Nuclear structure Vol II, (Benjamin, New York, 1975).

\bibitem{thskin2} V. G. Soloviev, Theory of complex nuclei, (Pergamon, Oxford, 1976). 

\bibitem{thskin3} P. Ring and P. Schuck, The nuclear many-body problem, (Springer, Berlin 1980). 

\bibitem{thskin4} S. Cwiok, J. Dudek, W. Nazarewicz, J. Skalski, and T. Werner, Comput. Phys. Commun. 46, 379 (1987). 

\bibitem{thskin5} S. G. Nilsson, I. Ragnarsson, Shapes and shells in nuclear structure, (Cambridge University Press, Cambridge, 1995). 

\bibitem{thskin6} Y. Aboussir, J. M. Pearson, and A. K. Dutta, At. Data Nucl. Data Tabl. 61, 127 (1995). 

\bibitem{thskin7} A. Baran, Z. Lojewski, K. Sieja, and M. Kowal, Phys. Rev. C72, 044310 (2005).
 
\bibitem{thskin8} M. Kowal, P. Jachimowicz, and A. Sobiczewski, Phys. Rev. C82, 014303 (2010).

\bibitem{pomorski1} J. F. Berger, K. Pomorski, Phys. Rev. Lett. 85, 30 (2000). 

\bibitem{pomorski2} A. Dobrowolski, K. Pomorski, and J. Bartel, Phys. Rev. C 65, 041306(R) (2002).

\bibitem{skin1} A. Trzcinska, et al., Phys. Rev. Lett. 87, 082501 (2001). 

\bibitem{skin2}A. Trzcinska, Acta Phys. Polonica B41, 311 (2010).

\end{thebibliography}
\end{document}